\documentstyle[aps,twocolumn,epsfig]{revtex}

\input epsf.sty

\begin{document}

\title{Inhibition of spontaneous emission in Fermi gases}

\author{Th. Busch,
J.R. Anglin,
J.I. Cirac,
and P. Zoller}

\address{Institut f\"ur Theoretische Physik, Universit\"at Innsbruck, 
A--6020 Innsbruck, Austria}

\newcommand{\vk}{\vec{k}}
\newcommand{\vke}{\vec{k}_1}
\newcommand{\vkp}{\vec{k}^\prime}
\newcommand{\vkpe}{\vec{k}^\prime_1}
\newcommand{\vl}{\vec{l}}
\newcommand{\vle}{\vec{l}_1}
\newcommand{\vlp}{\vec{l}^\prime}
\newcommand{\vlpe}{\vec{l}^\prime_1}

\maketitle

\begin{abstract}
Fermi inhibition is a quantum statistical analogue for the inhibition of
spontaneous emission by an excited atom in a cavity. This is achieved
when the relevant motional states are already occupied by a cloud of
cold atoms in the internal ground state. We exhibit non-trivial effects
at finite temperature and in anisotropic traps, and briefly consider a
possible experimental realization.
\end{abstract}
\date{5 May 1998}
\narrowtext

%%%%%%%%%%%%%%%%%%%%%%%%%%%%%%%%%%%%%%%%%%%%%%%%%%%%%%%%%%%%%%%%%%%%
%%%%%%%%%% Introduction
%%%%%%%%%%%%%%%%%%%%%%%%%%%%%%%%%%%%%%%%%%%%%%%%%%%%%%%%%%%%%%%%%%%%

The inhibition and enhancement of spontaneous emission of an atom by embedding
it in a cavity has been one of the most striking predictions in the field of
cavity QED \cite{Purcell,Kleppner}. Several experiments have and are currently
being carried out to test this prediction. The explanation of this phenomenon
is very simple: according to Fermi's golden rule, the spontaneous emission
rate $\Gamma$ is proportional to the density of modes of the electro-magnetic
field available at the atomic transition frequency; the cavity modifies the
mode structure, and therefore this rate. In particular, if all the normal
modes of the cavity are far off resonance with respect to the atomic
transition frequency, there are no modes available where a photon can be
emitted, and therefore spontaneous emission is inhibited.

In this paper we analyze how the spontaneous emission rate $\Gamma$ of
an atom can be inhibited by using quantum statistical principles. The
idea is to surround a fermionic atom in an internal excited state
$|e\rangle$ by a gas formed by identical particles in an internal ground
state $|g\rangle$. According to Fermi's golden rule, $\Gamma$ depends on
the occupation number of the different motional states to which the atom
can be transferred after spontaneous emission. In particular, if they
are already occupied then the atom cannot decay. This can be viewed as
well as a simple consequence of Pauli's exclusion principle. Thus, in
analogy to cavity QED one can say that spontaneous emission can be
inhibited by reducing the motional ``modes'' available to the excited
atom. 

From a qualitative point of view, the inhibition of spontaneous emission
can be easily understood. Consider a set of $N$ fermions in the state
$|g\rangle$ and one fermion in an excited state $|e\rangle$ that can
only decay into $|g\rangle$. Assume, for simplicity, that all the atoms
are trapped in a harmonic potential at a given temperature. In this
problem there are three energies scales which determine the properties
of $\Gamma$. First, $k_BT$, the thermal energy of the particles. Second,
the Fermi energy $E_F$ of the ground state atoms, which is the maximum
energy of the fermions at zero temperature. Finally, the recoil energy
$E_R=\hbar^2k_0^2/2m$ (where $k_0$ is the wavevector corresponding to
the transition $|e\rangle \rightarrow |g\rangle$ and $m$ is the atom
mass), which is the averaged motional energy acquired by the excited
atom in the emission process. Typically, the motional states with
energies $E\le E_F- k_BT$ are occupied, whereas those with $E\ge E_F+k_BT$
are basically available. Thus, if the initial motional energy
of the excited atom plus the one gained in the emission process $\sim
k_BT + E_R \le E_F -k_BT$, spontaneous emission will be inhibited.
As the temperature decreases or the Fermi energy increases
(equivalently, the number of ground state atoms increases) this
phenomenon becomes more pronounced. A quantitative analysis of this
phenomenon, on the other hand, is much more complicated since in order
to evaluate the results given by Fermi's golden rule one needs to
perform heavy numerical calculations. In this paper we have performed
these numerical calculations, as well as obtained simple analytical
formulas
for some interesting regimes. In particular, we have studied
the inhibition of spontaneous emission of photons along different
spatial directions both for isotropic and cylindrically symmetric traps
as a function of the temperature, Fermi energy and anisotropy parameter.
We have also analyzed a situation which we think it is closer to the
foreseeable experimental possibilities.

%%%%%%%%%%%%%%%%%%%%%%%%%%%%%%%%%%%%%%%%%%%%%%%%%%%%%%%%%%%%%%%%%%%%%
%%%%%%%%% Section I: The basic phenomenon
%%%%%%%%%%%%%%%%%%%%%%%%%%%%%%%%%%%%%%%%%%%%%%%%%%%%%%%%%%%%%%%%%%%%%

We consider an excited atom confined in a harmonic potential at some
given temperature $T$. Let us denote by $\Gamma_0(\Omega)d\Omega$ the
spontaneous emission rate of photons in the solid angle $d\Omega$ around
the direction defined by $\Omega$ for a single atom, and by
$\Gamma(\Omega)d\Omega$ the corresponding value in the presence of $N$
ground state fermions confined in the same potential at the same
temperature $T$. We are interested in the ``modification factor''
$M_f(\Omega) \equiv \Gamma(\Omega) /\Gamma_0(\Omega)$ which gives us
information about the modification of spontaneous emission along a given
direction $\Omega$. Using Fermi's golden rule we obtain
\begin{equation}
 \label{eq:inhibition}
  M_f(\Omega)=
  \sum_{\vec{m},\vec{n}=0}^\infty P_{\vec{m}}(1-F_{\vec{n}})
  |\langle \vec{n}|e^{-i\vec{k}(\Omega)\cdot\hat{\vec{r}}} |\vec{m}\rangle|^2
\end{equation}
Here, $F_{\vec{n}}=(e^{(\hbar\omega/k_BT)(\vec{\nu}\cdot\vec{n}-
\mu)}+1)^{-1}$ are the Fermi-Dirac and $P_{\vec{m}}=P_0
e^{-(\hbar\omega/k_BT)\vec{\nu}\cdot\vec{m}}$ the Boltzmann distribution
for the ground and excited state atoms; $|\vec{n}\rangle =
|n_x\rangle |n_y\rangle |n_z\rangle$ denote the eigenstates of the
harmonic oscillator with frequencies
$(\omega_x,\omega_y,\omega_z)\equiv\omega\vec{\nu}$ along the principal
axis \cite{FG}. 
The chemical potential $\mu$ and the number of atoms $N$ can be
expressed in terms of $E_F$ and $\omega$ in the usual way. Finally,
$\vec k(\Omega)/k_0$ is a unit vector along the direction given by
$\Omega$. The interpretation of this formula is very simple. The
probability density that the excited atom is transferred from $|\vec
m\rangle$ to $|\vec n\rangle$ by emitting a photon along the direction
$\Omega$ is $|\langle \vec{n}|e^{-i\vec{k}(\Omega)\cdot\hat{\vec{r}}}
|\vec{m}\rangle|^2$. This matrix elements has to be multiplied by the
probability that the initial state and the final state are occupied
$[P_{\vec m}$ and $(1-F_{\vec n})]$ and summed over. Note that, in order
to determine $M_f$, one has to perform six infinite summations for each
value of $\Omega$, which in general cannot be evaluated even using a
supercomputer. In the following we will consider that
$\omega_z/\lambda=\omega_x=\omega_y$, in which case
$M_f(\Omega)=M_f(\theta)$ only depends on the polar angle $\theta$ and
the problem becomes tractable numerically. Furthermore, we will also
derive some analytical approximations to this formula.

\begin{center}
 \begin{figure}
 \epsfig{file=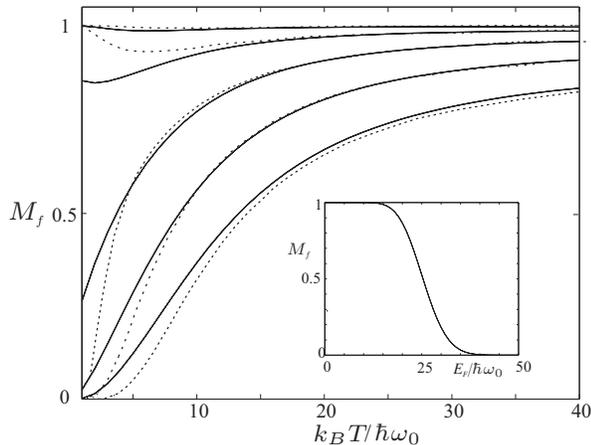,width=.9\linewidth} 
 \caption{
   $M_f$ as a function of temperature, for a trap with
   $E_R=25\hbar\omega$. The solid curves represent Fermi energies ranging in steps
   of ten from $E_F=10\hbar\omega$ (top) to $E_F=50\hbar\omega$ (bottom); the dotted curves
   are the semi-classical approximation.  The inset shows
   $M_f$ at $T=0$ as a function of Fermi energy.}
 \label{fig:inhib_iso}
 \end{figure}
\end{center}

To begin we consider an isotropic trap ($\lambda=1$). In this case,
using the spherical symmetry one can show that the emission rate becomes
independent of the photon direction and Eq.~(\ref{eq:inhibition})
reduces to
\begin{eqnarray}
 \label{eq:emission_iso}
 M_f=\frac{\Gamma}{\Gamma_0} &=&
     \sum_{n,m=0}^\infty P_m(1-F_n)\times\nonumber\\
     &&\sum_{l=0}^{\min[n,m]} 
     (l+1)|\langle n-l|e^{ik_0\hat{x}} |m-l \rangle|^2\;,
\end{eqnarray}
where $\Gamma_0$ and $\Gamma$ are the spontaneous emission rates in free
space and in the presence of quantum statistical effects, respectively.
Now, the states $|n\rangle$ denote one dimensional eigenstates of the
harmonic oscillator and $P_m$ and $F_n$ are the Boltzmann
and Fermi distributions at a given energy $n\hbar\omega$. These sums
can be performed numerically; the results are shown in Fig.\
\ref{fig:inhib_iso} as a function of temperature and Fermi energy. One
can see clearly that, as predicted by the simple arguments given above,
Fermi inhibition is strongest at low temperatures and large Fermi
energies, but does persist at relatively higher temperatures. 

At zero temperature the Fermi-Dirac distribution becomes a step
function, and the sums in Eq.~(\ref{eq:emission_iso}) may be evaluated
explicitly. The result is an incomplete Gamma function
$M_f=\Gamma[E_F/(\hbar\omega), E_R/(\hbar\omega)]$, whose integral
representation expresses the basic competition between Fermi energy and
recoil energy \cite{AS}
We have plotted this result in the inset of Fig.\ \ref{fig:inhib_iso}. 
As long as the Fermi energy is much smaller than the
recoil energy, the spontaneous emission rate of the atom is not
substantially different from the free case; this is because the matrix
element in Eq.~(\ref{eq:emission_iso}) is concentrated around $n-m=\pm
(E_R/\hbar\omega)$ (partial conservation of momentum due to the finite
strength of the confining potential). So Fermi inhibition at zero
temperature is negligible as long as levels around $E_R/(\hbar\omega)$
are still unoccupied. Once the Fermi sphere expands beyond these levels,
however, the probability of emission drops sharply, since to emit an
optical photon and move into an available motional state then requires
an improbably large impulse from the trapping potential.

At high temperatures, the sharp edge of the Fermi-Dirac distribution
erodes; eventually this means that we can analyze
Eq.~(\ref{eq:emission_iso}) semi--classically. The matrix element for
the $m \to n$ transition is concentrated around the values satisfying
$\hbar\omega(n-m) = \pm E_R$ with width $\delta E\sim 2(\hbar\omega
E_R)^{1/2}\sqrt{2m}$, where $m\hbar\omega$ is typically $\le 2k_BT$
(otherwise $P_m\simeq 0$). On the other hand, the Fermi-Dirac
distribution changes appreciably only at energy scales $k_BT$, i.e,
$F_n \simeq F_{n'}$ for $\hbar\omega|n-n'|\ll k_BT$.
Thus, if $\delta E \ll k_BT$, i.e. $E_R \ll k_BT$ we can replace $F_n
\to F_{m\pm E_R/(\hbar\omega)}$ in Eq.~(\ref{eq:inhibition}). The sum
over $n$ then equals one; and at high temperature we can also
approximate the sums over $m$ by integrals, obtaining
\begin{equation}
 \label{eq:inhib_semicl}
  M_f=\frac{\Gamma}{\Gamma_0}\doteq
   e^{\beta(\eta^2-E_F)}\log(1+e^{-\beta(\eta^2-E_F)})
\end{equation}
As can be seen from Fig.~\ref{fig:inhib_iso}, this approximation is
actually quite good even for moderately high temperatures. As $T\to
\infty$ the semi-classical result becomes exact, and shows that Fermi
inhibition disappears in the classical limit, when the chance of any
particular level being occupied becomes infinitesimal. By combining this
observation with the fact that for $E_F < E_R$ Fermi inhibition requires
temperatures sufficient to populate the levels near $E_R$, we can explain
why the upper curves in Fig.~\ref{fig:inhib_iso} actually show local
minima at finite temperatures.

%%%%%%%%%%%%%%%%%%%%%%%%%%%%%%%%%%%%%%%%%%%%%%%%%%%%%%%%%%%%%%%%%%%%%%%%%
%%%%%%%%%% Section II: The anisotropic case
%%%%%%%%%%%%%%%%%%%%%%%%%%%%%%%%%%%%%%%%%%%%%%%%%%%%%%%%%%%%%%%%%%%%%%%%%

We now consider the anisotropic case, $\lambda\not=1$. It is remarkable that
for an atom in free space spontaneous emission does not depend at all on the
shape of the trap \cite{CCT}. The presence of ground state fermions in an
anisotropic trap changes this fact: the spontaneous emission rate does depend
on the polar angle $\theta$ between the emission direction and the trap
$z$-axis. We emphasize that this is a specific consequence of quantum
statistics.

Returning to Eq.~(\ref{eq:inhibition}) for $\lambda \not=1$, we can
again begin by considering its simplified form at zero temperature.
Because of the remaining cylindrical symmetry of our trap, sums over
the three components of $\vec{n}$ reduce trivially to sums over
$n_x$ and $n_z$ only.  We find
\begin{eqnarray}
  \label{eq:radial}
  M_f(\theta)&=&
  e^{-\eta^2(\sin^2\theta+\lambda^{-1}\cos^2\theta)}\times\nonumber\\
  &&\sum_{n_x=0}^\infty 
  \sum_{n_z=\tilde n}^\infty
  \frac{\eta^{2(n_x+n_z)}\lambda^{-n_z}}{n_x!n_z!}
  [\sin\theta]^{2n_x}[\cos\theta]^{2n_z}\;,
\end{eqnarray}
where $\tilde{n}\equiv\max(0,{E_F-Nn_x\over\lambda})$, and we have used
$\eta^2=E_R/(\hbar\omega)$, where $\eta$ is the so--called Lamb--Dicke
parameter. 

\begin{figure}
 \centering
  \epsfig{file=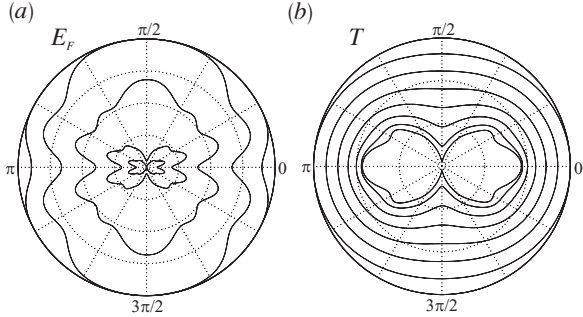,width=.9\linewidth}
 \caption{$M_f(\theta)$ in a trap with $E_R=25\hbar\omega$ and
   $\lambda=11$, for $(a)$ $T=0$, for $E_F/(\hbar\omega)$ ranging from $1$ 
   (outermost) to $45$ (innermost) in increments of 11; and $(b)$ 
   $E_F=45\hbar\omega$ for $T/(\hbar\omega)$ ranging
   from 0.5 (innermost) to 3.5 (outermost) in increments of 0.5.  The
   radial scale is linear, and the same for all curves; the radius of the
   circular frames is one.}
 \label{fig:radial}
\end{figure}

The results of evaluating numerically Eq.\ (\ref{eq:radial}) are
displayed in Fig.\ 2. Examining these results reveals, in addition to
fine structure for which no simple, qualitative explanation is apparent,
a rather surprising general feature. As $E_F$ rises from zero, Fermi
inhibition is initially strongest in the direction in which the trap is
stiffer; but at high $E_F$, it is emission in the soft direction which
is most inhibited. One might think to explain the greater initial
inhibition in the stiff direction by reference to the lower density of
states into which the emitting atom must recoil, if it recoils in the
stiff direction. This tempting explanation is incorrect, however,
because the larger trap force in the stiff direction allows greater
violation of momentum conservation, and consequently the range of energy
levels into which the atom may recoil is also greater. It is actually
trivial to show that the greater width of $|\langle
n|e^{ik\hat{x}}|m\rangle|^2$ exactly compensates for the lower density
of states, if all final states $|n\rangle$ are available (which is why
emission by a single atom is independent of the trap shape). And this
reveals the true explanation for the crossover from soft to stiff
directions. While $|\langle n|e^{ik\hat{x}}|m\rangle|^2$ is peaked at
the same energy $E_R$ regardless of $\lambda$, the wider distribution
for the stiff direction is first to be encroached on by the rising Fermi
level, which prevents recoil into occupied states. The wider
distribution is also last to be completely submerged as $E_F$ rises past
$E_R$, so for large $E_F$ spontaneous emission is preferentially in the
stiff direction. And the crossover point is once again $E_F=E_R$. 

While Fig.~2 presents results for a prolate (`cigar-shaped') trap, so
that `the soft direction' is actually the equatorial plane, to describe
an oblate (`pancake-shaped') trap one need merely rotate the circular
plots by ninety degrees. One piece of curious structure in $M_f(\theta)$
as a function of $E_F$ which is not shown in Fig. 2(a) is that if
$\lambda$ is an integer, $M_f(0)$ and $M_f(\pi)$ are 
$\lambda$-fold degenerate as functions of integer $E_F/(\hbar\omega)$.
If $\lambda$ is the inverse of an integer, $M_f(\pm\pi/2)$ are similarly
degenerate. 
Fig. 2(b) shows that at finite temperatures $M_f(\theta)$ becomes more
isotropic. Indeed, at very high temperature complete isotropy is
restored, as Pauli exclusion becomes insignificant in the classical
limit. Therefore the occurrence of the spatial anisotropy of the spectrum can also be
used signaling the onset of the degenerate regime in the cooling process
of the gas.

%%%%%%%%%%%%%%%%%%%%%%%%%%%%%%%%%%%%%%%%%%%%%%%%%%%%%%%%%%%%%%%%%%%%%%%%%%
%%%%%%%%  Section III: Putting in the excited atom with lasers
%%%%%%%%%%%%%%%%%%%%%%%%%%%%%%%%%%%%%%%%%%%%%%%%%%%%%%%%%%%%%%%%%%%%%%%%%%

So far we have assumed that the one excited atom had initially been
brought into the trap without any disturbance of the equilibrium
population of ground state atoms. But let us now consider the
experimentally more straightforward scenario in which the excited atom
has been created by applying a laser pulse to a ground state cloud. To
ensure that no more than one excited atom is involved, we suppose a weak
pulse. On the other hand, we will assume that the pulse is short
compared to the internal and external dynamics of the atoms, so that we
can consider it instantaneous. Taking the excited atom from the ground
state cloud introduces two significant changes. Firstly,
even at zero temperature there is a significant probability of the
excited atom having energy above $E_F$, because it may have been excited
from a motional state near the Fermi level. And secondly, the excitation
of one atom leaves behind a hole in the Fermi sea, into which the atom
can always recoil.

Recalculating the rate of spontaneous emission in an isotropic trap, from the 
initial state with one atom excited from the cloud as we have described, we 
obtain
\begin{equation}
\label{ff}
  \tilde{M_f}(\Omega)=1+\frac{1}{N}
         \left[ |\sum_{\vl} \langle n_{\vl}\rangle C_{\vl\vl}(\Omega)|^2-
         \sum_{\vl\vlp}\langle n_{\vl}\rangle\langle n_{\vlp}\rangle
         |C_{\vl\vlp}(\Omega)|^2 \right]\;,
\end{equation}
where 
$\tilde{M_f}(\Omega)=\Gamma(\Omega)/[N\Omega_R^2t_0^2\Gamma_0(\Omega)]$ is the emission
rate in the direction $\Omega$ given that an atom has indeed been
excited. $\Omega_R$ is the Rabi frequency and $t_0$ the duration of the
applied laser pulse. The $C_{\vl\vlp}(\Omega)$ are the transition matrix
elements between the states $|\vlp\rangle$ and $|\vl\rangle$. The
direction $\Omega$ is specified by the polar an azimuthal angles
$(\theta,\phi)$ with respect to the propagation direction of the
exciting laser pulse. Obviously, the results for the isotropic trap
depend only on $\theta$.

The first term within the large brackets in Eq.\ (\ref{ff}) represents
processes in which an atom is excited by the laser pulse, then spontaneously
emits a photon and recoils back into the `hole' from which it came.  Because
the final state of the gas is the same no matter which atom undergoes this
process, absorption and re-emission by all $N$ atoms interfere constructively,
and so this term is of order $N^2$; but since the atom recoils back into the
state from which it was kicked by the laser pulse, the re-emitted photon is
overwhelmingly likely to be in the same direction as the pulse \cite{note}.
The second term in the brackets of Eq. (6) describes Fermi-inhibited
spontaneous emission just as we have discussed in the first part of this
Letter, but with the more complicated initial state for the excited atom.  It
is this term that is of interest, but since it is only proportional to $N$, it
is dominated by the first term within a cone around the forward direction.
Evaluating the two terms for small angles from the laser direction, one finds
the width of this cone to be $|\delta\theta|\sim(\eta N^{1/6})^{-1}\sim 1/(k_0
L)$, where $L$ denotes the atomic cloud size (see Fig.\ 3).  This is the
familiar results of diffraction of light by a transparent object of size $L$.

We have plotted in Fig.\ 3 $\tilde{M_f}(\theta)$ as function of
$\theta$. For $\theta>\delta\theta$, Fermi inhibition is indeed
significant, though it is nowhere close to the near-total suppression
found in the cases where the excited atom is initially thermal. This is
because there is always a significant fraction of the ground state atoms
within two recoil energies of $E_F$, and these can be excited and
spontaneously emit without much hindrance.

\begin{center}
\begin{figure}
 \epsfig{file=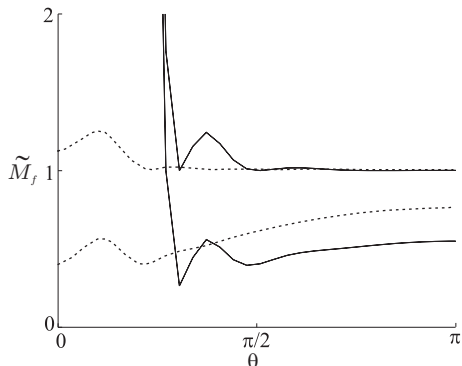,width=0.7\linewidth}
  \caption{$\tilde{M_f}$ at $T=0$ for a trap with $E_R=\hbar\omega$ and
    $E_F=10\hbar\omega$.  The solid line corresponds to Eq.\ \ref{ff} and the
    dashed line to the case where forward scattering is suppressed by a Raman
    pulse excitation. In both
    cases the above lying curve is missing the Fermi-inhibition term (the
    last one in Eq. \ref{ff}).}
 \label{fig:laser}
\end{figure}
\end{center}

To see Fermi inhibition without simply waiting for rare emissions
outside the forward cone, one possibility is to make the exciting pulse
a Raman pulse, so that the recoil momentum when the atom emits one
photon is different from its recoil on absorbing the two Raman photons.
By this means one can ensure that recoil back into the hole the excited
atom came from is strongly suppressed by momentum conservation, and the
order $N^2$ term in Eq.\ (6) can be essentially eliminated. This effect
is shown by the dotted curves in Fig.\ 3. We emphasize, however, that
the Raman laser should involve a dipole forbidden transition in order for
the $|e\rangle\to |g\rangle$ to be allowed, which makes the experiment
more challenging.

Finally, we would like to mention that in the calculations presented here we
have neglected atom--atom collisions, which is a good approximation since
s--wave scattering is suppressed for spin polarized Fermions. On the other
hand, Fermi's golden rule neglects the effects of dipole--dipole interactions
and reabsorption, and therefore our model assumes that the atomic gas is
optically thin, meaning that the probability of reabsorbtion of the emitted
photon $P_{reab}=\frac{\sigma_E}{4\pi}\langle\frac{1}{r_{12}^2} \rangle$
\cite{reab} is negligible. Some straightforward calculations show that this
implies the condition $0.15E_f^2/(E_R\hbar\omega)\ll 1$, which for stiff traps
will still allow significant Fermi inhibition.

In conclusion, we have analyzed the possibility of observing Fermi
inhibition of spontaneous emission in an atomic gas. We have derived
simple analytical formulas for the modification factor $M_f$
and compared with numerical results for both isotropic and cylindrically
symmetric traps. We have also studied the situation in which an
atom is excited by a weak short laser.

This work was supported by the European Union under the TMR Network
ERBFMRX-CT96-0002 and by the Austrian Fond zur F\"orderung der
wissenschaftlichen Forschung. We thank W. Ketterle for pointing us out that 
Fermi inhibition was already predicted in \cite{Pritchard}.

% --------------------------------------------------------


\begin{references}
\bibitem{note}
In the limit where the time-scale for this 
forward re-emission is as short as the laser pulse itself, it is of course 
necessary to re-interpret forward spontaneous emission as coherent forward 
scattering.  Re-summing these terms to all orders would then reveal how the 
propagation of the pulse through the cloud is modified by the index of 
refraction of the gas.  But we will assume that the pulse is so short compared 
to $1/\Gamma$ that we are far from this limit.
\end{references}
\end{document}